\documentclass[twocolumn]{aastex62}

\usepackage{graphicx}   
\usepackage{amsmath}    
\usepackage{mathrsfs}    
\usepackage{bm}
\usepackage{amssymb}    
\usepackage{diagbox}    
\usepackage{subfigure}
\usepackage{enumerate}
\usepackage{float}
\usepackage{ulem}
\usepackage{url}
\usepackage{lineno}
\shorttitle{Maximum accreted mass of recycled pulsars}
\shortauthors{Z. Li et al.}

\begin{document}

\title{The maximum accreted mass of recycled pulsars }

\correspondingauthor{Zhenwei Li}
\email{lizw@ynao.ac.cn}

\author[0000-0002-1421-4427]{Zhenwei Li}
\affiliation{Yunnan Observatories, Chinese Academy of Sciences, Kunming, 650011, People's Republic of China}
\affiliation{Key Laboratory for the Structure and Evolution of Celestial Objects, Chinese Academy of Science, People's Republic of China}
\affiliation{University of the Chinese Academy of Science, Yuquan Road 19, Shijingshan Block, 100049, Beijing, People's Republic of China}


\author{Xuefei Chen}

\affiliation{Yunnan Observatories, Chinese Academy of Sciences, Kunming, 650011, People's Republic of China}
\affiliation{Key Laboratory for the Structure and Evolution of Celestial Objects, Chinese Academy of Science, People's Republic of China}
\affiliation{Center for Astronomical Mega-Science, Chinese Academy of Science, 20A Datun Road, Chaoyang District, Beijing 100012, People's Republic of China} 
\nocollaboration

\author{Hai-Liang Chen}

\affiliation{Yunnan Observatories, Chinese Academy of Sciences, Kunming, 650011, People's Republic of China}
\affiliation{Key Laboratory for the Structure and Evolution of Celestial Objects, Chinese Academy of Science, People's Republic of China}
\nocollaboration

\author{Zhanwen Han}

\affiliation{Yunnan Observatories, Chinese Academy of Sciences, Kunming, 650011, People's Republic of China}
\affiliation{Key Laboratory for the Structure and Evolution of Celestial Objects, Chinese Academy of Science, People's Republic of China}
\affiliation{Center for Astronomical Mega-Science, Chinese Academy of Science, 20A Datun Road, Chaoyang District, Beijing 100012, People's Republic of China} 
\nocollaboration




\begin{abstract}

  The maximum mass of neutron stars (NSs) is of great importance for constraining equations of state of NSs and understanding the mass gap between NSs and stellar-mass black holes. NSs in X-ray binaries would increase in mass by accreting material from their companions (known as recycling process), and the uncertainties in the accretion process give challenge to study the NS mass at birth. {In this work, we investigate the NS accreted mass with considering the effect of NS spin evolution and give the maximum accreted mass for NSs in the recycling process. By exploring a series of binary evolution calculations, we obtain the final NS mass and the maximum accreted mass for a given birth mass of NS and a mass transfer efficiency. Our results show that the NSs can accrete relatively more material for binary systems with the donor masses in the range of $1.8\sim 2.4M_\odot$, the NSs accrete relatively more mass when the remnant WD mass is in the range of $\sim 0.25-0.30M_\odot$, and the maximum accreted mass is positively correlated with the initial NS mass. For a $1.4M_\odot$ NS at birth with a moderate mass transfer efficiency of 0.3, the maximum accreted mass could be $0.27M_\odot$. The results can be used to {estimate} the minimum birth mass for systems with massive NSs in observations. }

\end{abstract}




\section{Introduction}
\label{sec:1}

A neutron star (NS) is the remnant of a massive star. NSs are supposed to be produced from electron capture supernovae and core-collapse supernovae (\citealt{nomoto1984,nomoto1987,burrows1995,takahashi2013,wangb2020}; see \citealt{woosley2002} for a review). The mass of NS at birth is strongly dependent on the supernova explosion processes (\citealt{timmes1996,tauris2015}; and references therein). Theoretically, according to the different density profiles of NS, the maximum masses can range from $\sim1.5$ to $\sim 2.8M_\odot$ (\citealt{rikovska2007,read2009,goriely2010,potekhin2013,kojo2015}; and \citealt{ozel2016} for a recent review). However, most NSs are observed with mass less than $2.0M_\odot$. Recent pulsar radio timing and X-ray observations found several NSs with mass beyond $\sim 2M_\odot$, e.g. PSR J1600-3053 with a NS mass of $2.3^{+0.17}_{-0.15}M_\odot$ \citep{arzoumanian2018}, PSR J2215+5135 with a NS mass of $2.28^{+0.10}_{-0.09}M_\odot$ \citep{kandel2020}, PSR J1959+2048 with a NS mass of $2.18\pm 0.09M_\odot$ \citep{kandel2020}, PSR J0740+6620 (hereafter J0740) with a NS mass of $2.072^{+0.067}_{-0.066}M_\odot$ \citep{cromartie2020,riley2021}, and PSR J0348+0432 (hereafter J0348) with a NS mass of $2.01\pm 0.04M_\odot$ \citep{antoniadis2013}. The detection of gravitational-wave event GW190814 also suggests that there is a possibility for the existence of NS with a mass {around $2.6M_\odot$ \citep{GW190814}}. The massive NSs in observation is important for inferring the NS mass distribution \citep{valentim2011,kiziltan2013,alsing2018,shaod2020}, and constraining the NS equation of state \citep{lim2020,godzieba2021}. 

Many NSs are in binary systems, including X-ray binaries (NSs are in the accretion phase), double NSs, and NS+white dwarf binaries, NS+black hole binaries, etc. \citep{ozel2016, LIGO2021}. Most of NSs can accrete material from the companions and have been spun up (\citealt{alpar1982,radhakrishnan1982,bhattacharya1991}; see also \citealt{tauris2012}). There are many uncertainties on the recycling phase, e.g. accretion efficiency (ratio of NS accreted mass to the transferred mass from donor), accretion disk instability, propeller effect \citep{antoniadis2012,antoniadis2016,paradijs1996,romanova2018}, etc. Therefore, it is unclear how much material can be accreted by NS in that phase. {In an extreme case, if we only consider the spin-up process, the spin period of NS is inversely correlated to the NS accreted mass. For example, the NS can be spun-up to $10\;\rm ms$ and $1\;\rm ms$ by accretion of $\sim 0.01M_\odot$ and $\sim 0.22M_\odot$, respectively \citep{tauris2012}. However, due to the existence of the spin-down process during the mass transfer phase, NS may accrete more mass for a given recycled spin period \citep{liuw2011}. The exact amount of mass accreted by NS during the recycling process is affected by the detailed treatment of the mass transfer, and is of great importance for constraining the birth mass of NS \citep{tauris2011,cognard2017}. }

In this work, we attempt to find out the maximum accreted mass of NS during the recycling processes by modelling the binary evolution with NS companions. {In many of previous works, the NS mass accretion is only limited by the Eddington rate, e.g. \citet{tauris1999,podsiadlowski2002,linj2011,van2019b}. Such a treatment simplifies the accretion process, and likely overestimates the accreted masses of NSs. Here we consider the effect of spin evolution of NS during the accretion processes in addition to the limit of Eddington rate, as done in \citet{tauris2011,liuw2011}. In this case, the propeller effect may occur and prevent mass accretion.} 

The paper is structured as follows. We present the model inputs and methods in Section 2, and the results are given in Section 3. The main uncertainties in our simulations are discussed in Section 4. Finally, we give the summary and conclusion in Section 5.

\section{Model inputs and methods}
\label{sec:2}
\subsection{Binary evolution code}
\label{subsec:2.1}
{Since there are many uncertainties during supernovae and NS birth, we start our study from a NS with a zero-age main sequence (ZAMS) star as a companion. The companion may overfill its Roche lobe and transfer material to the NS. If the mass transfer is dynamically unstable, the NS will be involved in the envelope of the companion and the binary will enter into common envelope evolution process. The common envelope evolution is complicated and whether the NS can accrete material during the common envelope phase is under debated \citep{ivanova2013,macleod2015,holgado2018}. We therefore do not consider NS accretion in this case in our study. If the mass transfer is dynamically stable, NS will accrete mass from the companion via stable mass transfer, we focus on this case here.} 

The detailed binary evolution calculations are done with the state-of-the-art stellar evolution code Modules for Experiments in Stellar Astrophysics (MESA, version 9575, \citealt{paxton2011,paxton2013,paxton2015}). For convenience, the neutron star (NS) is taken as a point mass. For the donor star, the initial element abundances of Population I stars, i.e. metallicity $Z=0.02$ are adopted. The hydrogen mass fraction is given by $X=0.76-3Z$ \citep{pols1998}. The mixing-length parameter is set to be $\alpha_{\rm MLT}=1.9$. The mass transfer rate is calculated with Ritter scheme (\citealt{ritter1988}), that is, 
\begin{eqnarray}
  \dot{M} \propto \frac{R^3_{\rm RL,d}}{M_{\rm d}}\exp\left(\frac{R_{\rm d}-R_{\rm RL,d}}{H_{\rm p}}\right),
\end{eqnarray}
where $R_{\rm d}$ and $R_{\rm RL,d}$ are the donor radius and its Roche lobe radius, {$M_{\rm d}$ is the donor mass}, and $H_{\rm p}$ is the pressure scale height.  

{The initial NS, $M_{\rm NS,i}$, ranges from 1.10 to 2.2 $M_\odot$, where the NS mass has a step of $0.2M_\odot$ from $1.4$ to $2.2M_\odot$, the choice of $1.10 M_\odot$ is intended to cover the minimum NS in the observtions (i.e. $1.174\pm 0.004M_\odot$ for the companion of PSR J0453+1559; \citealt{martinez2015}), and $1.25M_\odot$ is regarded as the mean mass of NS from electron capture supernovae \citep{schwab2010}.} {We assume that all transferred material from the donor flows to the NS, then some ($\beta_{\rm mt}$) is lost from the binary, some ($\delta_{\rm mt}$) forms circumbinary (CB) disk (see more details in Section 2.2). The remaining ($1-\beta_{\rm mt}-\delta_{\rm mt}$) is defined as mass transfer efficiency $f_{\rm mt}$. The value of $f_{\rm mt}$ is quite uncertain, and is set from $0.1$ to $0.9$ with a step of $0.1$. Note that the accreted mass of NS is also limited by the Eddington rate and inefficient accretion stage (see more details in Section 2.3), and the real accretion efficiency\footnote{{The definition of accretion efficiency here is a little different from that in \citet{antoniadis2016}. \citet{antoniadis2016} suggest that the accretion efficiency should be less than $0.2$ according to observations. However, it is noted that the accretion efficiency defined in their work is an average value, i.e. the fraction of NS accreted mass to the lost mass from the donor. While in this work, the accretion efficiency is defined in every time interval during the mass transfer phase. Due to the existences of Eddington limit and inefficient accretion, even if a high mass transfer efficiency is adopted, the value of average accretion efficiency is comparable with that in \citet{antoniadis2016}. For example, as shown in the left panel of Figure \ref{fig:2}, the mass transfer efficiency is $0.9$, the NS accretes $0.23M_\odot$ from a donor with mass of $2.0M_\odot$, and the remnant WD mass is $0.28M_\odot$, then the average accretion efficiency is $0.13$.}} is lower than $f_{\rm mt}$.} With a given $M_{\rm NS,i}$ and $f_{\rm mt}$, the initial donor masses range from 1.0 to 3.6 $M_\odot$ with a variable step size\footnote{{The reason of the variable size for $M_{\rm d,i}$ and $P_{\rm orb,i}$ is for the sake of computation cost. We find the accreted mass is relatively small for binaries with massive donors and wide periods, as shown in Section 3.3. Therefore, the step size of the initial parameters for these binaries are widen properly.} }, i.e. the donor mass has a step of $\triangle M_{\rm d}=0.2M_\odot$ for $M_{\rm d} \leq 2.4\;M_\odot$, and $\triangle M_{\rm d}=0.4M_\odot$ for $2.4<M_{\rm d}\leq 3.6M_\odot$. The initial orbital period ranges from $0.7$ to $2.0\;\rm d$ with a step of $0.05\;\rm d$, and from $2$ to $20\;\rm d$ in a step of $\Delta \log_{10}(P_{\rm orb,i}/{\rm d})=0.025$. For binary systems in wider orbits, the donors generally enter into the red giant branch at the onset of mass transfer. The mass transfer rate is significantly larger than the Eddington rate, and most of the envelope masses of the donors are lost from the system (see more details in Section \ref{subsec:3.1}). For a given $f_{\rm mt}$ and $M_{\rm NS,i}$, we will obtain the maximum increased mass of NS. {The evolution is stopped as the evolutionary age reaches 14 Gyr, but we mainly focus on the mass transfer stage, and the termination of mass transfer is defined as $\dot{M}\leq 10^{-12}M_\odot \;\rm yr^{-1}$ \citep{chenx2017}.} 

In this work, we mainly consider the binaries that evolve into detached NS + WD systems. The case of accreting pulsars with very low-mass non-degenerate companions, e.g. redbacks and black widows \citep{roberts2013,chenh2013}, are not included in our simulations. With the loss of the orbital angular momentum due to the gravitational wave radiation, the WD will fill its Roche lobe and transfer mass to the NS. The stability of mass transfer processes is still under debated \citep{haaften2012,bobrick2017,yus2021}, therefore, we ignore the cases of NS accreting mass from the WDs.

\subsection{Angular momentum loss}
\label{subsec:2.2}
We consider three types of angular momentum loss mechanism, which are gravitational wave radiation, magnetic braking, and mass loss, respectively. 

The orbital angular momentum carried away by the gravitational wave radiation can be calculated as \citep{landau} 
\begin{eqnarray}
  \dot{J}_{\rm GW} = -\frac{32G^{7/2}M_{\rm NS}^2M_{\rm d}^2(M_{\rm NS}+M_{\rm d})^{1/2}}{5c^5a^{7/2}},
\end{eqnarray}
where $M_{\rm NS}$ and $M_{\rm d}$ denote the NS and the donor star mass, $a$ is the semi-major axis of the orbit, $c$ is the speed of light in vacuum, and $G$ is the gravitational constant.  

The angular momentum loss because of magnetic braking is calculated from the formula \citep{rappaport1983}
\begin{eqnarray}
  \dot{J}_{\rm MB} = -3.8\times10^{-30}M_{\rm d}R^{\gamma_{\rm MB}}_{\rm d}\Omega^3\;\rm dyn\; cm,
\end{eqnarray}
{where $\gamma_{\rm MB}$ is the magnetic braking index, and is set to be $4$ according to the standard magnetic braking prescription\footnote{{The higher $\gamma_{\rm MB}$ means the stronger angular momentum loss caused by magnetic braking. However, the orbital evolution during the mass transfer stage, which is the main stage we concern, is mainly dominated by mass loss \citep{istrate2014a}. The varying of $\gamma_{\rm MB}$  will have little impact on our results.}}\citep{chenh2013,van2019a}}, $R_{\rm d}$ is the radius of the donor, $\Omega$ is the spin angular velocity, which equals to the orbital angular velocity $\omega_{\rm orb}$ as tidal synchronization is assumed. The magnetic braking effect can be neglected if the convective envelope becomes too thin. In this work, we switch on magnetic braking when the convective envelope fraction is larger than 0.01, as is done in \citet{chenx2017}. 

During mass transfer, we also consider the angular momentum extracted by the CB disk. The angular momentum loss rate under this torque can be expressed as (\citealt{spruit2001}; see also \citealt{shaoy2012,chenh2013}) 
\begin{eqnarray}
  \dot{J}_{\rm CB} = \gamma\left(\frac{2\pi a^2}{P_{\rm orb}}\right)\delta_{\rm mt} \dot{M}_{\rm d}\frac{t}{t_{\rm vi}}^{1/3},
\end{eqnarray}
where $\dot{M}_{\rm d}$ is the mass transfer rate,{$\gamma^2$ is the scale factor, and given by $r_{\rm i}/a$, where $r_{\rm i}$ is the inner radius of the disk, $a$ is the binary separation,} $t$ is the time since mass transfer begins, and $t_{\rm vi}$ is the viscous timescale at the inner edge of the disk, which is defined by $t_{\rm vi}=2\gamma^3P_{\rm orb}/3\pi\alpha\beta^2$, $\alpha$ is the viscosity parameter \citep{shakura1973}, $\beta$ is the ratio of the scale height to the radius of the disk. {Based on the observed results, we set $\gamma^2=1.7$ \citep{muno2006}, $\alpha=0.01, \beta=0.03$ \citep{belle2004}, and $\delta_{\rm mt}=3\times 10^{-4}$ \citep{taam2003}. }

The extra material that leaves the systems is assumed to take away the specific angular momentum of NS. Then the angular momentum loss due to mass loss is 
\begin{eqnarray}
  \dot{J}_{\rm ML} = -(1-f_{\rm mt}-\delta_{\rm mt})|\dot{M}_{\rm d}|\left(\frac{M_{\rm NS}}{M_{\rm NS}+M_{\rm d}}\right)^2\frac{2\pi a^2}{P_{\rm orb}}.
\end{eqnarray}

\subsection{Mass accumulation process of neutron star}
\label{subsec:2.3}

\begin{figure*}
    \centering
    \includegraphics[width=\textwidth]{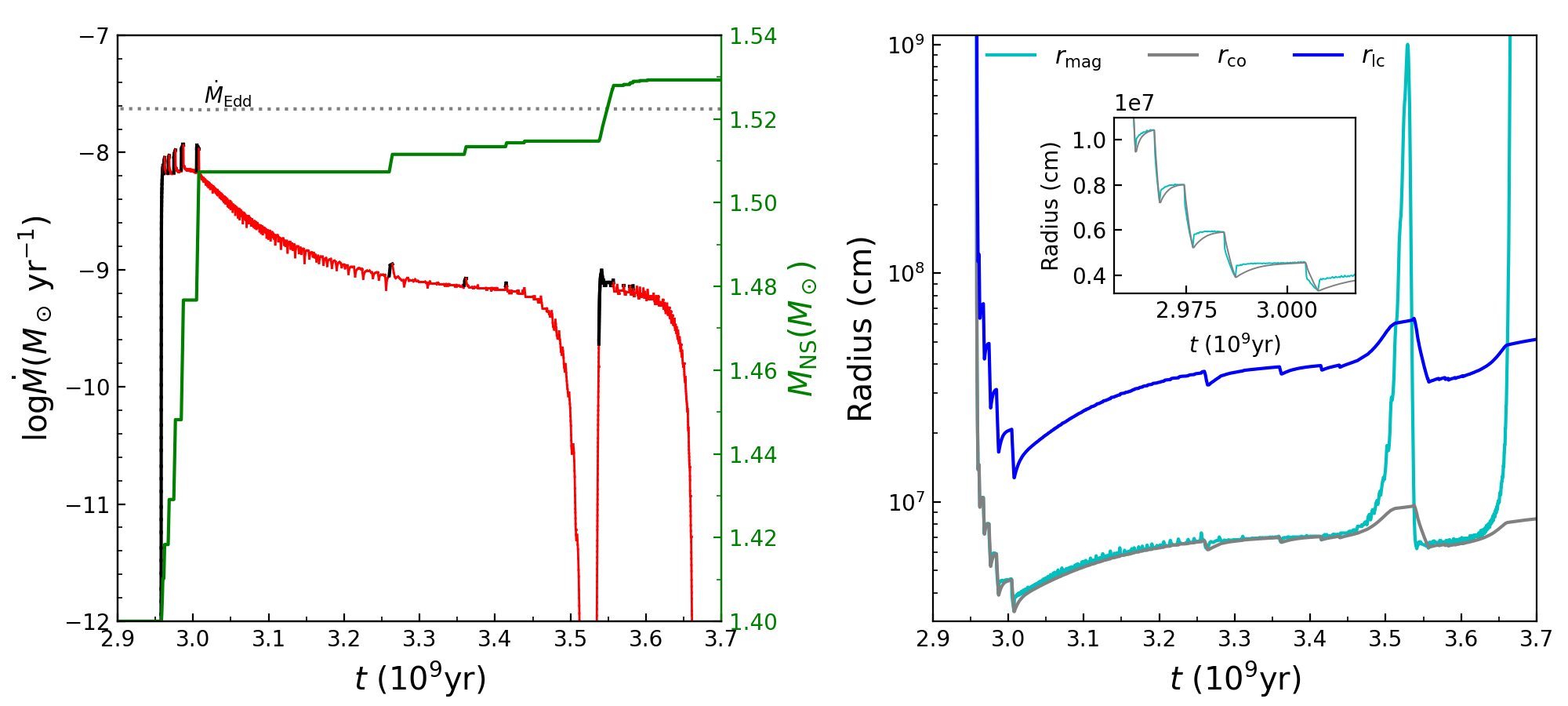}
    \caption{Left panel: The mass transfer rate and NS mass vs. star age for $M_{\rm d,i}=1.4M_\odot,P_{\rm orb,i}=1.6\;\rm d,M_{\rm NS,i}=1.4M_\odot$ and $f_{\rm mt}=0.9$. The mass transfer rate and NS mass are shown in black and green, respectively. The propeller effects are shown in the red lines. The Eddington rate is shown in grey dotted line. Right panel: The comparisons among $r_{\rm mag}$ (cyan line), $r_{\rm co}$ (grey line) and $r_{\rm lc}$ (blue line) are shown. The repeated accretion processes from $2.96$ to $3.02\;\rm Gyr$ are shown in the inset. See more details in the text.}
    \label{fig:1}
\end{figure*}

{Firstly, the accretion of the NS is limited by the Eddington accretion limit}\footnote{{There are several observation evidences for the existence of super-Eddington accretion in NS. For example, the recent detection of ultra-luminosity X-ray (ULX) source NGC 7793 P13 are supposed to be a NS accreting at super-Eddington rates \citep{israel2017}. Theory suggests that the NS in such system have a strong surface magnetic field ($\gtrsim 10^{14}\;\rm G$), and highly super-Eddington emission from NSs are likely rare events (\citealt{Tsygankov2016}; and \citealt{kaaret2017} for a review). The NSs concerned in this work have a relatively small magnetic field ($\leq 10^{12}\;\rm G$), and would not have significant super-Eddington accretion.}}, $\dot{M}_{\rm Edd}$, 
\begin{eqnarray}
  \dot{M}_{\rm Edd} &=& 3.6\times 10^{-8}\left(\frac{M_{\rm NS}}{1.4M_\odot}\right)\left(\frac{0.1}{GM_{\rm NS}/R_{\rm NS}c^2}\right)\nonumber\\
  && \times \left(\frac{1.7}{1+X}\right)M_\odot\rm yr^{-1},
\end{eqnarray}
where $R_{\rm NS}$ is the NS radius, and can be approximately expressed as a simple non-relativistic degenerate Fermi-gas polytrope: $R_{\rm NS}=15(M_{\rm NS}/M_\odot)^{-1/3}$ \citep{tauris2012}. Combining the limit of Eddington accretion rate, the accretion rate of the NS is 
\begin{eqnarray}
  \dot{M}_{\rm acc} = {\rm min}(-f_{\rm mt}\dot{M}_{\rm d},\dot{M}_{\rm Edd}).
\end{eqnarray}

{Secondly, the spin evolution of the NS is considered in addition to the limit of Eddington rate, which leads to an inefficient accretion during the mass transfer stage, as described below.} 

We define the magnetosphere radius, $r_{\rm mag}$ of the NS at which the ram pressure of the accreted material equals to the magnetic pressure in the magnetosphere \citep{lamb1973,ghosh1979a,ghosh1979b,liuw2011}, that is 
\begin{equation}
  r_{\rm mag} = 1.8\times 10^8 \left(\frac{B_{\rm s}}{10^{12}\rm G}\right)^{4/7}\left(\frac{\dot{M}_{\rm acc}}{10^{-8}M_\odot\;\rm yr^{-1}}\right)^{-2/7}\;\rm cm, 
\end{equation}
where $B_{\rm s}$ is the surface magnetic field of the NS, and $\dot{M}_{\rm in}$ is the mass inflow rate. The evolution of magnetic field during accretion process is described as \citep{shibazaki1989,wijers1997} 
\begin{equation}
  B_{\rm s} = \frac{B_{\rm i}}{1+\Delta M_{\rm acc}/m_{\rm B}}, 
\end{equation}
where $B_{\rm i}$ is the initial magnetic field of the NS, and is set to be $10^{12}\;\rm G$, $\Delta M_{\rm acc}$ is the accreted mass of the NS, {$m_{\rm B}$ is the mass constant for the field decay, and is set to be $10^{-4}M_\odot$ according to the observations \citep{shibazaki1989,zhangc2006,wangj2011}.} If the magnetosphere radius is less than the corotation radius, the infalling material can be accreted onto the NS surface. The corotation radius is defined as \citep{liuw2011,romanova2018} 
\begin{equation}
  r_{\rm co} = 1.5\times 10^8\left(\frac{M_{\rm NS}}{M_\odot}\right)^{1/3}P_{\rm spin}^{2/3} \rm\; cm,
\end{equation}
where $P_{\rm spin}$ is the spin period of NS in units of second. The spin-up torque during the accretion process is given by 
\begin{equation}
  \dot{J}_{\rm acc} = \dot{M}_{\rm acc}\sqrt{GM_{\rm NS}R_{\rm NS}}.
\end{equation}

With the spin-up of NS, $r_{\rm mag}$ will be greater than $r_{\rm co}$ at some point. In this situation, the centrifugal barrier at $r_{\rm mag}$ prevents the infalling material from being accreted by the NS. This process is known as propeller effect. The spin evolution during the propeller phase is calculated approximately by \citep{alpar2001} 
\begin{equation}
  \dot{J}_{\rm prop} = \dot{M}_{\rm acc}r^2_{\rm mag}[\Omega - \Omega_{\rm K}(r_{\rm mag})].
\end{equation}
Moreover, if $r_{\rm mag}$ is larger than the light cylinder radius $r_{\rm lc}$, where $r_{\rm lc} = c/\Omega  = 48\;{\rm km}(P_{\rm spin}/1\;\rm ms)$, the NS spins too fast to allow the infalling material penetrate the light cylinder and the NS appears as a radio pulsar. 

There is a maximum spin frequency for NS, i.e. Keplerian frequency, $f_{\rm K}(M_{\rm NS})$. The accreted material is supposed to be ejected if the spin frequency of NS equals to $f_{\rm K}(M_{\rm NS})$. Here $f_{\rm K}(M_{\rm NS})\simeq C\; {\rm kHz}\; (M_{\rm NS}/M_\odot)^{1/2}(R_{\rm NS}/10\;\rm km)^{-3/2}$, where $C$ is a fitted parameter and is set to be 1.15 \citep{haensel2009}. Since NS is an extremely compact object, the moment of inertia of NS should be calculated with general relativity effects and the specific equation-of-state of NS considered \citep{arnett1977}. For convenience, we take the NS as a point mass, and adopt $I=10^{45}\;\rm g\;cm^2$ for all kinds of NSs. The influence of $I$ on the NS accreted mass will be discussed in Section 4.2. 

{From what has been introduced above, the accretion efficiency , $\epsilon_{\rm acc}$, is given by} 
\begin{equation}
  \epsilon_{\rm acc} = \left\{
  \begin{aligned}
    &{\rm min}(f_{\rm mt},|\dot{M}_{\rm Edd}/\dot{M}_{\rm d}|),& {\rm Accretion\; phase} ; \\
    &0,& {\rm Inefficient\; accretion,} 
  \end{aligned}
  \right.
\end{equation}
{where the inefficient accretion cases includes that the propeller effects occur, NS is in the radio phase and NS spins at the Keplerian frequency.}

\section{Binary evolution Results}
\label{sec:3}

\subsection{Evolutionary examples}
\label{subsec:3.1}
The increase of NS mass is connected with the mass transfer rate during binary evolution, as shown in the left panel of Figure \ref{fig:1}, where the binary initially contains a $1.4M_\odot$ NS and a $1.4M_\odot$ donor. The mass accretion rate of NS can be easily obtained by using equation (7), and is not shown for clarity. At the early stage of mass transfer, the mass transfer is on a thermal timescale. Therefore, the NS mass can increase rapidly. The propeller effect starts to work after the NS accretes a small part of masses. At this moment, the magnetosphere radius is larger than corotation radius, as shown in the inset of right panel (where cyan line is above the grey line). The matter is unable to be accreted onto the NS due to the centrifugal force exerted by the magnetosphere\footnote{The magneto-hydrodynamic simulations (e.g. \citealt{romanova2018}) suggest that a considerable part of matter can penetrate the magnetosphere and accreted by the NS in a weaker propeller regime. However, the specific propeller efficiency (characterize the relative amount of matter ejected from the binary) is rather uncertain and strongly depends on the assumptions (see more details in \citealt{romanova2018}). The scope of this work is to give a lower limit of NS accreted mass during recycling phase, therefore, the accretion in the weaker propeller regime is ignored. }. Meanwhile, the centrifugal barrier exerts a propeller spin-down torque on the NS in that phase. As the spin period increases, the magnetosphere radius can be less than the corotation radius at some point (where the grey line is above the cyan line in the inset of right panel), resulting in the repeated accretion processes at the early mass transfer phase. And the NS can accrete about $0.15M_\odot$ during that phase. After the initial thermal timescale mass transfer, the mass transfer rate decreases due to the radius expansion of star driven by the nuclear burning \citep{podsiadlowski2002}, and the NS enters into a long-term propeller phase. The sudden decrease of mass transfer around $3.52\;\rm Gyr$ is due to the discontinuity of the composition gradient during the first dredge-up stage \citep{tauris1999,istrate2016a}. There is still enough envelope material for burning after the dredge-up, the donor star expands again and a subsequent mass transfer occurs \citep{jiak2014,lizw2019}. It is noted that the NS can only accrete very little material in that epoch and most transferred material is accreted during the thermal timescale mass transfer stage.

\begin{figure*}
    \centering
    \includegraphics[width=\textwidth]{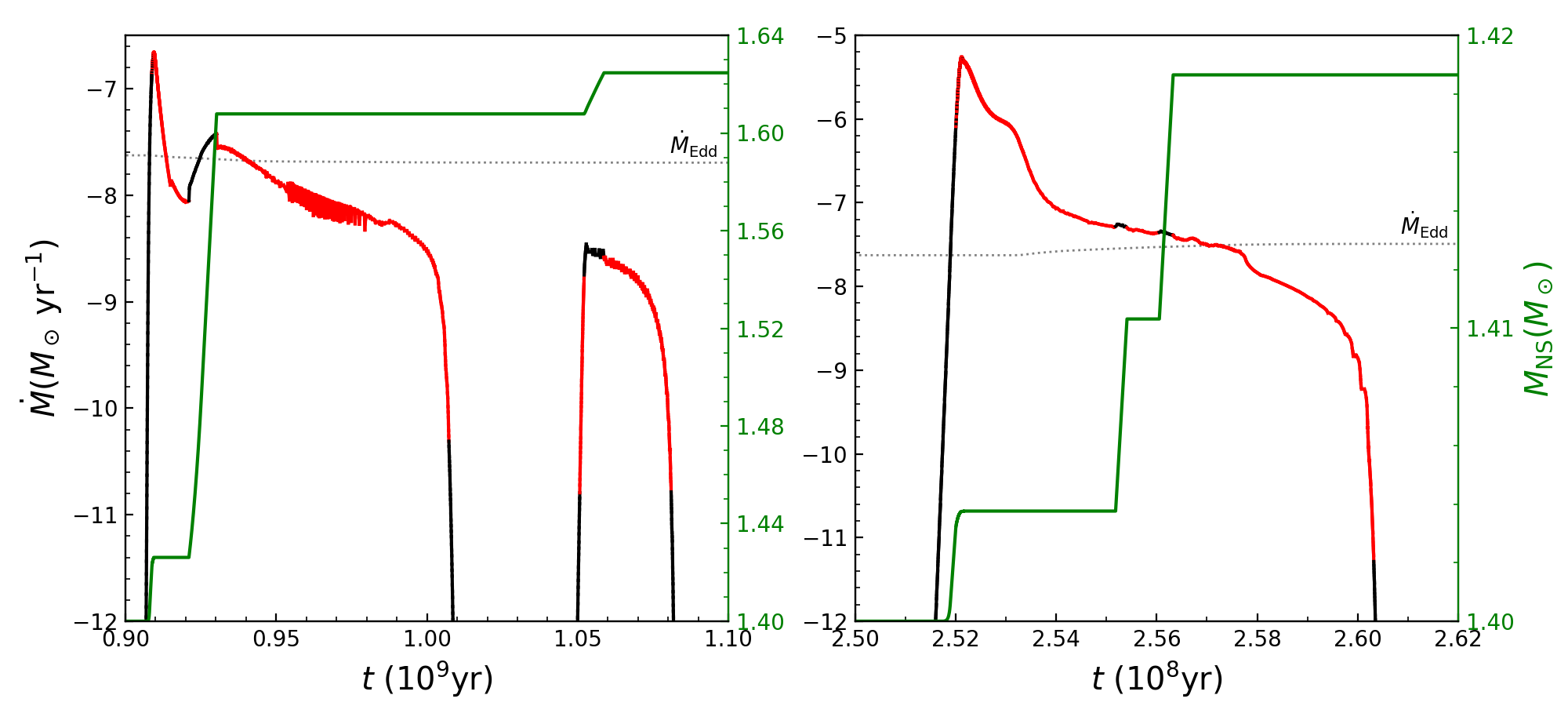}
    \caption{Similar to the left panel in Figure \ref{fig:1}, but for $M_{\rm d,i}=2.0M_\odot,P_{\rm orb,i}=1.3\;\rm d$ and $M_{\rm d,i}=3.2M_\odot,P_{\rm orb,i}=1.6\;\rm d$ with NS mass of $1.4M_\odot$ and $f_{\rm mt}=0.9$ from left to right, respectively.}
    \label{fig:2}
\end{figure*}

\begin{figure*}
    \centering
    \includegraphics[width=\textwidth]{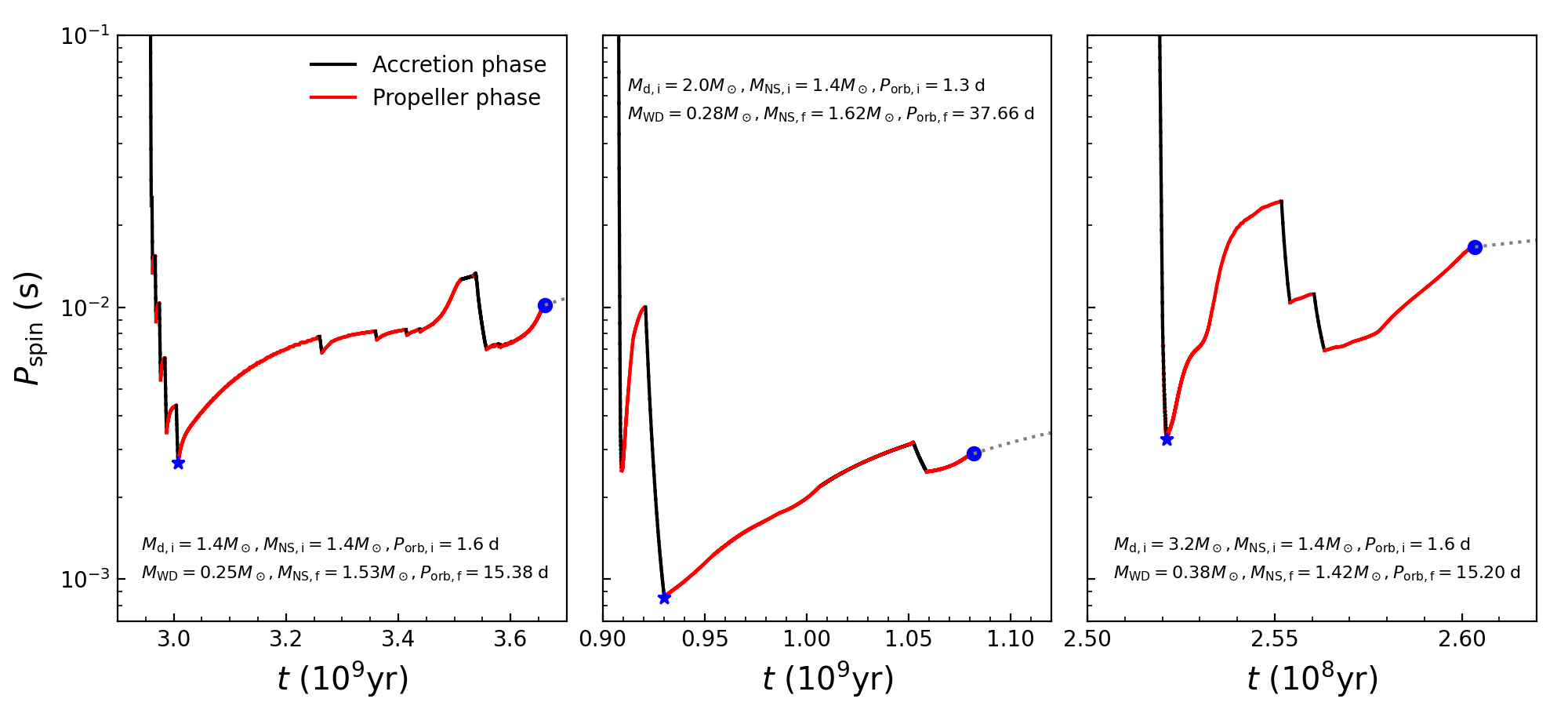}
    \caption{The spin evolution of the three examples in Figure \ref{fig:1} and \ref{fig:2}. The initial and final parameters are shown in each panel. The accretion and propeller phases are shown in solid black and red lines, and the blue stars and circles are for the minimum spin periods and spin periods at the end of mass transfer, respectively. The NSs spinning down due to magnetic dipole radiation after the mass transfer phase are shown in dotted lines. }
    \label{fig:3}
\end{figure*}

\begin{figure}
    \centering
    \includegraphics[width=\columnwidth]{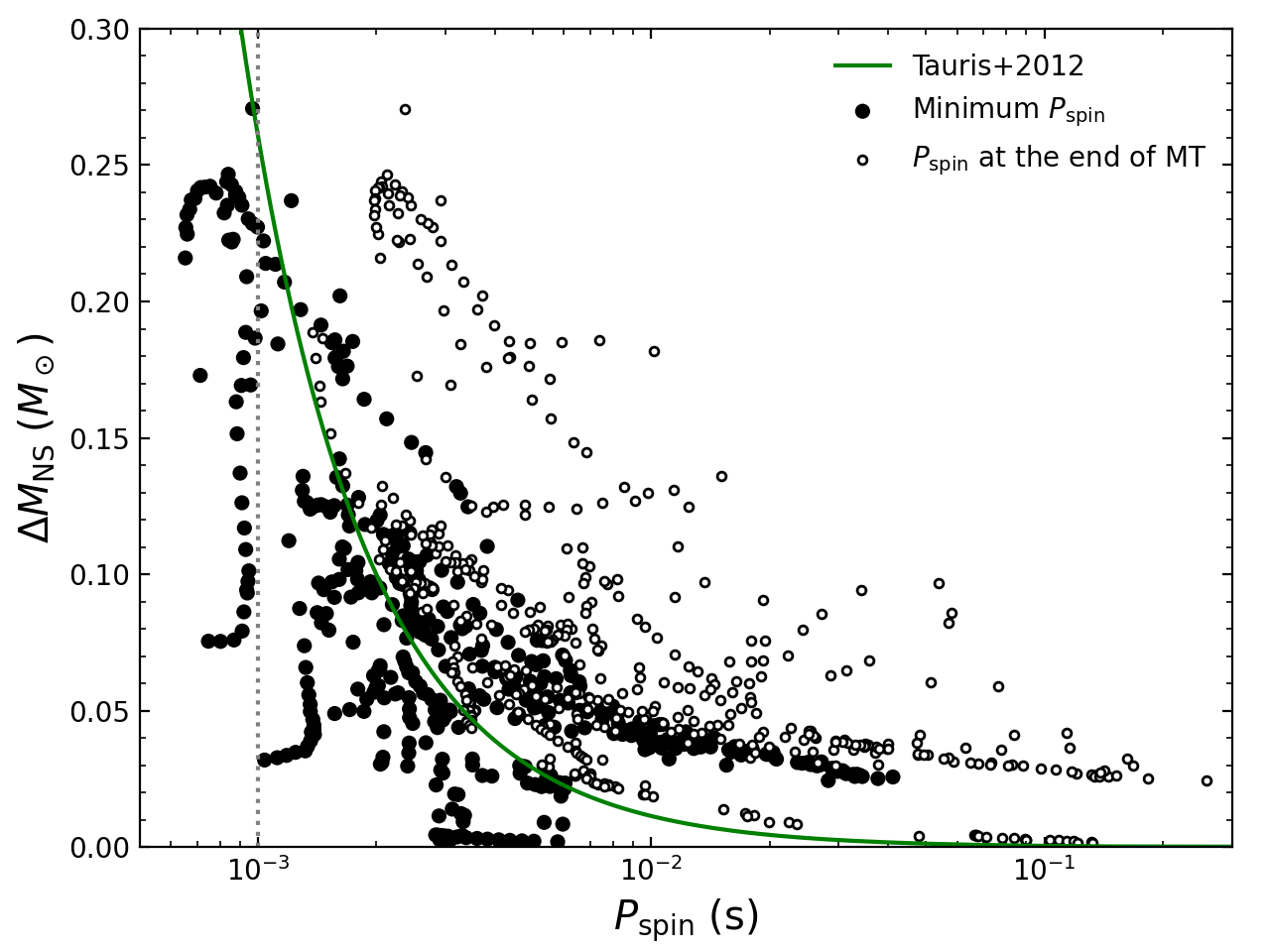}
    \caption{The spin period of NS vs. the accreted mass for $f_{\rm mt} = 0.3$ and $M_{\rm NS,i}=1.4M_\odot$. The open circles are for the spin period at the end of mass transfer, and the solid circles are for the minimum spin period during the accretion processes. The accreted mass as a function of a given spin period (the initial NS mass is set to be $1.4M_\odot$) in \citet{tauris2012} is shown in green line for comparison.  
    }
    \label{fig:4}
\end{figure}

As shown in Figure \ref{fig:1}, how much mass can be accreted by NS during the recycling process strongly depends on the mass transfer process. For low-mass donors, e.g. $M_{\rm d,i}=1.4M_\odot$ in the {left panel of Figure \ref{fig:1}}, the mass transfer rate is always sub-Eddington. While for more massive donors, the mass transfer rate may exceed the Eddington limit, such as in the cases of Figure \ref{fig:2}. We see that the NS can accrete relatively more masses from intermediate-mass donors ($1.8\lesssim M_{\rm d,i}\lesssim 2.4M_\odot$, e.g. the case in the left panel of Figure \ref{fig:2}). The reason is that the main accretion process for NS occurs during the thermal timescale mass transfer phase. While for massive donors ($M_{\rm d,i}\gtrsim 2.4M_\odot$, e.g. the case in the right panel of Figure \ref{fig:2}), the thermal timescale mass transfer rate could be larger than the Eddington rate by several orders of magnitude. Therefore, a significant part of transferred mass will be ejected due to the Eddington limit \citep{podsiadlowski2002}. As a result, the NS would not accrete too much material from a low-mass donor ($M_{\rm d,i}\lesssim 1.8M_\odot$) due to the low thermal timescale mass transfer rate, and also cannot accrete too much material from a massive donor since most of the transferred material is ejected on account of the Eddington limit. 

\subsection{The spin evolution of NS}

{In Figure \ref{fig:3}, we present the spin evolution of NS with different initial parameters, as shown in the panels. Most material has been accreted by the NS during thermal timescale mass transfer, resulting in a rapid decrease of the spin period. The NS accreted masses from left to right are $0.13,0.22,0.02M_\odot$, respectively, and the NS generally rotates fast with a large accreted mass. In the middle panel, the minimum spin period of the NS is less than $1\;\rm ms$. However, no sub-millisecond pulsars have been discovered yet \citep{hessels2006,papitto2014,patruno2017,bassa2017,haskell2018}. The cause may be that the timescale of NS in the sub-millisecond stage is very short, about $10^7$ yr as shown in the middle panel. As a comparison, the timescale of NS with $P_{\rm spin}\lesssim 10\;\rm ms$ is about $2\times 10^{8}\;\rm yr$, which is twenty times larger than that of sub-millisecond stage. At the end of mass transfer, the spin periods (as shown in blue circles) are several times larger than the minimum spin period, and the NS subsequently spins down due to the magnetic dipole radiation \citep{tauris2012}. }

{The spin period of NS versus the accreted mass is shown in Figure \ref{fig:4}, where the open circles are for the spin period at the end of mass transfer, and the solid circles are for the minimum spin period during the accretion process. In general, the NS rotates fast for a large $\Delta M_{\rm NS}$, consistent with the theoretical result in \citet{tauris2012}. The reason of the dispersion around the theoretical curve is that we consider the specific mass transfer process during the accretion. We also found that some NSs may have sub-millisecond spin period, but they will spin down due to the propeller effects and the spin periods for the simulated samples at the termination of mass transfer are larger than $1\;\rm ms$. {Besides, at the end of mass transfer, the accreted mass is larger than that calculated in \citet{tauris2012} for a given recycled spin period (i.e. the open circles are above the green line). The reason is that the NS may accrete more mass and obtain a relatively shorter spin period, and then spins down to the given recycled spin period (as shown in Figure 3).}

\subsection{The mass increase of NS and the remnant WD mass}
\begin{figure}
    \centering
    \includegraphics[width=\columnwidth]{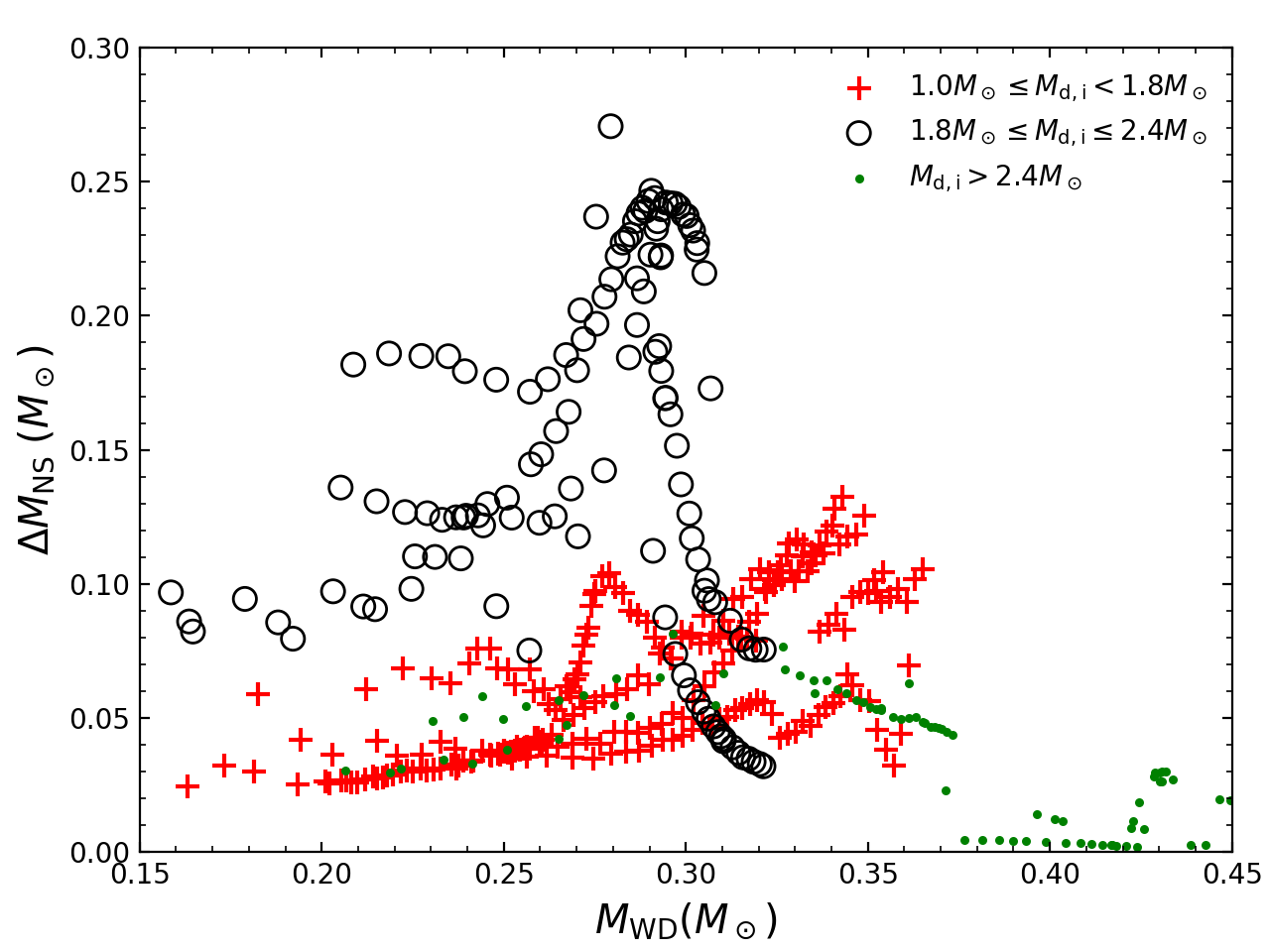}
    \caption{The relation between NS accreted mass and WD mass for different donors with $M_{\rm NS,i}=1.4M_\odot,f_{\rm mt}=0.3$. The symbols represent the donors in the different mass range, as indicated in the Figure. In general, the NS can accrete more material if the donor mass is in the range of $1.8-2.4M_\odot$. See more details in the main text.} 
    \label{fig:5}
\end{figure}

To find the maximum accreted mass of NS for a given $f_{\rm mt}$ and $M_{\rm NS,i}$, we present the relation between NS accreted mass and the remnant WD mass, as shown in Figure \ref{fig:5}, where $M_{\rm NS,i}=1.4M_\odot,f_{\rm mt}=0.3$. The symbols represent the donors in the different mass range, as indicated in the panels. It is clear that the NSs accrete relatively more masses as $M_{\rm d,i}$ ranges from $1.8M_\odot$ to $2.4M_\odot$. The reasons are that the mass loss caused by Eddington limit is not too much, and the NS can accrete relatively more material during the thermal timescale mass transfer phase, as discussed in Section \ref{subsec:3.1}.

The NSs can accrete relatively more masses for $M_{\rm WD}$ around $0.25-0.30M_\odot$. As discussed above, the main accretion process occurs at the early stage of mass transfer phase. {Binary with short orbital period generally leads to a low thermal timescale mass transfer rate and a small remnant WD mass. Therefore, we see that the NS accretes $0.1M_\odot$ at most with $M_{\rm WD}\simeq 0.15M_\odot$ in Figure \ref{fig:5}, which is lower than that with $M_{\rm WD}\simeq 0.25-0.30M_\odot$. For binaries with large orbital periods, most of the transferred material is lost due to the Eddington limit, and the NS accreted mass is less than $0.05M_\odot$ for $M_{\rm WD}$ around $0.40M_\odot$}. The fluctuation of NS accreted mass is ascribed to the propeller effect, which depends not only on the mass transfer rate, but also on the surface magnetic field of the NS and the corotation radius (see equations 8-10).

{Figure \ref{fig:5} shows a correlation between the NS accreted mass and the remnant WD mass. If we assume all NSs have similar birth masses in binary pulsars, there should be a correlation between the final NS masses and the WD masses. For example, the NSs could be more massive with $M_{\rm WD}$ in the range of $0.25-0.30M_\odot$. However, the relation becomes uncertain when the birth masses of NSs distribute in a large range, as found in observations \citep{rawls2011,kandel2020,faulkner2005,janssen2008,arzoumanian2018}. In a further work, we will explore the mass distribution of NS and its companion by combining the simulation results and binary population synthesis method, and try to find the correlation between NS mass and companion mass.} 

\subsection{The maximum accreted mass of NSs}

\begin{figure}
    \centering
    \includegraphics[width=\columnwidth]{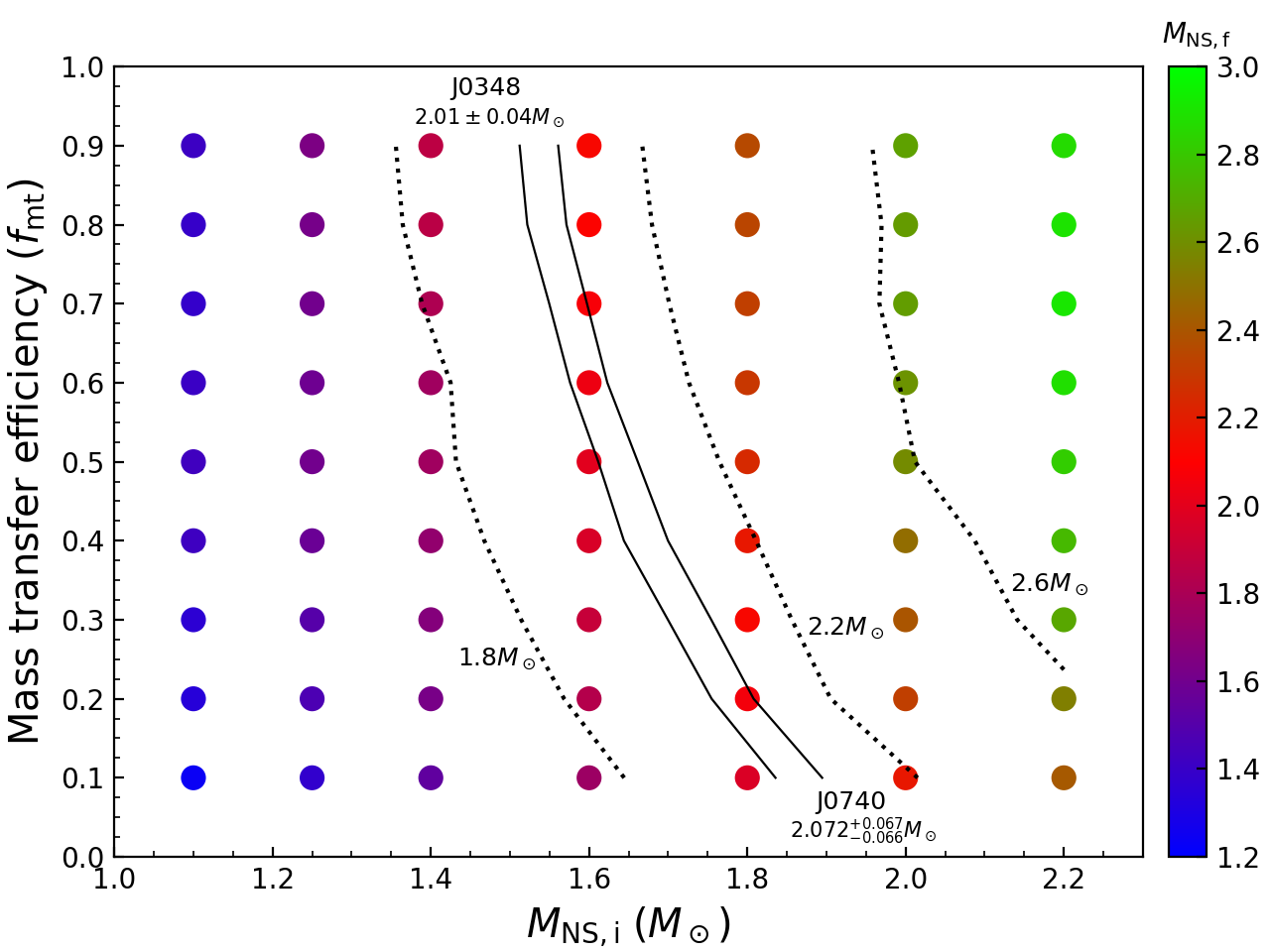}
    \vfill
    \includegraphics[width=\columnwidth]{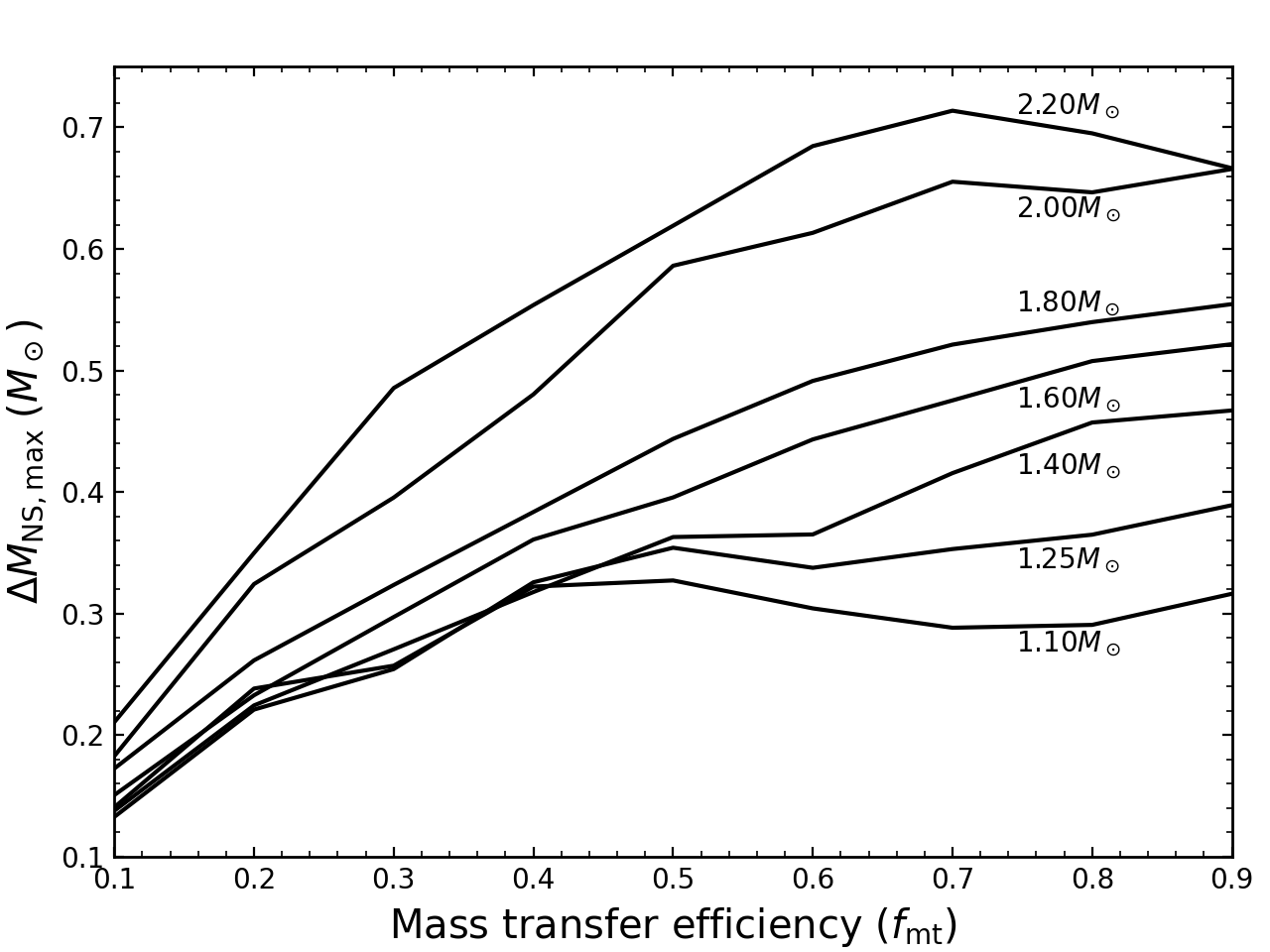}
    \caption{Upper panel: The maximum NS mass after the accretion with given mass transfer efficiencies and initial NS masses. The three dotted lines represent the final NS mass of $1.8$, $2.2$ and $2.6M_\odot$, respectively. And the two observed pulsars with masses larger than $2M_\odot$ are shown in black solid lines. Lower panel: The maximum accreted mass of NSs with different $f_{\rm mt}$ and initial NS masses. The initial NS masses from the bottom to top are $1.10,1.25,1.40,1.60,1.80,2.00,2.20M_\odot$, respectively.}
    \label{fig:6}
\end{figure}

{The NS accreted mass is strongly dependent on the initial binary parameters, and is hard to be determined. However, for a given $f_{\rm mt}$ and $M_{\rm NS,i}$, there is a maximum accreted mass in the simulations. For example, when $f_{\rm mt}=0.3$ and $M_{\rm NS,i}=1.4$, the maximum accreted mass of NS is about $0.27M_\odot$, as shown in Figure 5. By changing the values of $f_{\rm mt}$ and $M_{\rm NS,i}$, we may get the corresponding maximum accreted mass in a similar way. In the upper panel of Figure \ref{fig:6}, we present the maximum NS mass after the accretion with given mass transfer efficiencies and initial NS masses, where the final NS masses are shown in colors. For clarity, we plot several dotted lines to express a given final NS mass as noted in the figure. The curves are given by the linear interpolation between adjoining grids. The relations between NS maximum accreted mass and mass transfer efficiency for different initial NS mass are shown in the low panel. We see that the maximum accreted masses vary little for $f_{\rm mt}$ larger than $\sim0.5$ due to the existence of propeller effect. In general, massive NSs can accrete more material in comparison with low-mass NSs if other parameters are fixed. For example, a binary with a $1.25M_\odot$ NS, the NS can accrete about $0.39M_\odot$ mass for $f_{\rm mt} = 0.9$. However, for an initial NS mass of $2.2M_\odot$, the maximum accreted mass could be $\sim 0.66M_\odot$. The reason is that the massive NS has a relatively large corotation radius, resulting in more material captured by the NS.} 

{The two massive pulsars with He WD companions, i.e. PSR J0348 and J0740, are shown in solid lines in the upper panel. It is noted that the maximum accreted mass of NS is different for each $M_{\rm NS,i}$. If a moderate mass transfer efficiency of 0.3 is assumed, the minimum birth mass of NS can be obtained by the interpolation between the adjoining grids in the upper panel of Figure \ref{fig:6}. We find that the NS birth masses should be larger than $1.70$ and $1.75M_\odot$ for PSR J0348 and J0740, respectively. The results of minimum NS birth masses with different mass transfer efficiencies speculated for the two observed samples are presented in Table 1.} 

\startlongtable
\begin{deluxetable}{lcc}
  \tablecaption{The minimum NS birth masses ($M_{\rm NS,min}$) of J0348 ($2.01\pm0.04M_\odot$) and J0740 ($2.072^{+0.067}_{-0.066}M_\odot$) inferred from this work.}
\tablecolumns{6}
\tablenum{1}
\tablewidth{0pt}
\tablehead{
\colhead{$f_{\rm mt}$} &
\colhead{$M_{\rm NS,min} (M_\odot)$} &
\colhead{$M_{\rm NS,min} (M_\odot)$}\\
\colhead{} & 
\colhead{J0348} &
\colhead{J0740}
}
\startdata
0.1 & $1.84$ & $1.89$\\
0.2 & $1.75$ & $1.81$\\
0.3 & $1.70$ & $1.75$\\
0.4 & $1.64$ & $1.70$\\
0.5 & $1.61$ & $1.66$\\
0.6 & $1.57$ & $1.62$\\
0.7 & $1.55$ & $1.59$\\
0.8 & $1.52$ & $1.57$\\
0.9 & $1.51$ & $1.56$\\
\enddata
\end{deluxetable}


{Black widows and redbacks are one particular class of recycled pulsars that have been suggested to have significant accretion during the recycled processes \citep{roberts2013}. In these pulsar binaries, the very low-mass non-degenerate companions are irradiated and ablated by the NSs \citep{chenh2013}. However, the NSs in such binaries are still in the accretion phase, which are not considered in our simulations. In the recent observations, two of such binaries are found with massive NSs, i.e. a NS mass of $2.18\pm0.09M_\odot$ for PSR J1959+2048 and a NS mass of $2.28^{+0.10}_{-0.09}M_\odot$ for J2215+5135. Chen et al. (2013) studied the formation of this kind of pulsar binaries, and found that the low-mass companions can be produced from donors with mass around $1.0-1.2M_\odot$ by considering the evaporation effects. In the formation scenario of NS with low-mass companion, a degenerate core is not be developed at the onset of mass transfer. Therefore, the progenitor binaries are supposed to have short orbital period, which leads to a relatively lower mass transfer rate. As discussed in section 3.1, the NS mainly increases its mass during the thermal timescale mass transfer, and the accreted mass of NS in such binaries should not be larger than the maximum accreted mass calculated in this work.}

\section{Uncertainties in the simulations}
We consider the spin evolution of NS during the recycling processes of a NS. The main uncertainties in the simulations are the assumptions of initial magnetic field and the NS moment of inertia. In this section, We give a discussion about the influences of these two parameters. 
\subsection{The effect of initial NS magnetic field}
{Most NSs at birth may have magnetic field higher than $\sim 10^{13}\;\rm G$ \citep{haberl2007}. Before the onset of mass transfer, the magnetic field decays due to the Ohmic decay of electric currents located in the NS crust or core. Then the NS may have a magnetic field weaker than $10^{12} \;\rm G$ at the onset of mass transfer process \citep{aguilera2008,gullon2014,bransgrove2018}. }A weak pulsar magnetic field will lead to a small accretion radius (magnetosphere radius), and angular momentum can be effectively transferred to the NS, which results in a large spin-up rate \citep{longair}. In general, the propeller phase occurs ($r_{\rm mag}>r_{\rm co}$) slightly later for NS with a weak magnetic field, then more material could be accreted by the NS. 

We calculated the evolution of binaries with $M_{\rm NS,i}=1.4M_\odot$, $B_{\rm mag,i}=0.5\times 10^{12}\;\rm G$. For $f_{\rm mt}=0.9$, the NS accreted mass for models with small $B_{\rm mag,i}$ is higher than that for models with $B_{\rm mag,i}=10^{12}\;\rm G$ by about $0.22M_\odot$. With the decrease of mass transfer efficiency $f_{\rm mt}$, the influence of initial magnetic field correspondingly decreases. For example, the difference of maximum NS accreted mass between $B_{\rm mag,i} =0.5\times 10^{12}\;\rm G$ and $10^{12}\;\rm G$ for $f_{\rm mt}=0.1$ is about $0.025M_\odot$. {Therefore, the choice of the initial magnetic field has a limited effect for a small $f_{\rm mt}$, but has a significant effect for a large $f_{\rm mt}$.}

\subsection{The influence of NS moment of inertia}
The NS moment of inertia is hard to solve due to the uncertainties of the NS equation-of-state. In this work, we adopt a constant value of $I$ ($10^{45}\;\rm g\; cm^2$) in the calculations, which is almost the lower limit for NS. The true value of moment of inertia for massive NS may be larger than $3\times 10^{45}\;\rm g \; cm^2$ (\citealt{greif2020}, and references therein). Here we discuss the effect of moment of inertia on our results. 

A larger $I$ means the acceleration of the spin-up process during the accretion phase is small, which results in a long spin period during the accretion phase. According to equation (10), the corotation radius is larger than that for NS with a low-value $I$. As a result, the binary spends a relatively shorter time on propeller phase for NS with a larger $I$, and the NS can accrete more material during the accretion phase. To illustrate this issue, we additionally calculate the cases of $1.4M_\odot$ with $I=2\times 10^{45}\;\rm g\; cm^2$. For a large mass transfer efficiency of $f_{\rm mt}=0.9$, the NS accreted mass is about $0.1M_\odot$ greater than the case of $I=1\times 10^{45}\;\rm g\; cm^2$. Similarly, the influence of $I$ decreases when $f_{\rm mt}$ becomes small.

\section{Summary and Conclusion}
\label{sec:6}
In this work, we consider the spin evolution of NSs, and calculate the maximum accreted mass of NS in a binary system. Our main conclusions are summarized as follows.

(1) The accreted masses are strongly dependent on the initial donor mass and the remnant WD mass. In general, the NS can accrete relatively more material for donor mass in the range of $1.8M_\odot\sim2.4M_\odot$ than that of donors in the other mass range. The NS accretes relatively more mass when the remnant WD mass is in the range of $\sim 0.25-0.30M_\odot$. 

(2) The maximum accreted mass of NS is positively correlated to the initial NS mass. In other words, massive NSs can accrete more material than low-mass NSs with other initial parameters fixed. 

(3) With the consideration of spin evolution for the NS, the maximum NS accreted masses change little for mass transfer efficiency $f_{\rm mt}>0.5$ because of the propeller effects. 

{(4) The maximum accreted masses of NSs with different $f_{\rm mt}$ and $M_{\rm NS,i}$ are given. For example, for a NS with birth mass of $1.4M_\odot$, if we assume a moderate mass transfer efficiency of $0.3$, the NS can accrete $\sim 0.27M_\odot$ at most. In the extreme case, if $f_{\rm mt}=0.9$, the maximum accreted mass of NS is about $0.465M_\odot$. }

{(5) We analyze two massive pulsars with WD companions, i.e. J0348 ($M_{\rm NS}=2.01\pm0.04M_\odot$) and J0740 ($M_{\rm NS}=2.062^{+0.067}_{-0.066}M_\odot$). Both of them are supposed to experience recycling process during the mass transfer phase. If a moderate mass transfer efficiency of $0.3$ is adopted, the birth masses of NSs should be larger than 1.70 and 1.75$M_\odot$ for PSR J0348 and J0740, respectively. In an extreme case of $f_{\rm mt}=0.9$, the birth masses of NSs should be larger than $1.51$ and $1.56M_\odot$ for PSR J0348 and J0740, respectively.}

{The results addressed in this work can be used to estimate the minimum birth mass for the observed pulsars. It is hard to give a more constraint on the likely birth mass of NSs since the accreted mass is strongly dependent on the initial progenitor binary parameters. However, by comparison of the WD mass distribution for the NS+WD populations between observations and that of binary population synthesis could give some clues on this. For example, according to the relation between the WD mass distribution and the accreted mass distribution, we may estimate the likely accreted mass for the NS. Besides, the mass transfer efficiency has a significant effect on the NS accreted mass, which can be limited by the NS mass distribution.}

\section*{Acknowledgements}
The authors gratefully acknowledge the “PHOENIX Supercomputing Platform” jointly operated by the Binary Population Synthesis Group and the Stellar Astrophysics Group at Yunnan Observatories, Chinese Academy of Sciences. This work is partially supported by the Natural Science Foundation of China (Grant no. 11733008, 11521303, 11703081, 11422324), by the National Ten-thousand talents program, by Yunnan province (No. 2017HC018), by Youth Innovation Promotion Association of the Chinese Academy of Sciences (Grant no. 2018076) and the CAS light of West China Program. 
\software{MESA (v9575; \citealt{paxton2011,paxton2013,paxton2015})}

\bibliographystyle{./aasjournal}
\bibliography{./hns}

\include{table_information}
\include{table_KS}
\include{figure}

\end{document}